\documentclass[preprint,12pt]{elsarticle}

\usepackage{amssymb}
\usepackage{amsmath}
\usepackage{lineno}

\usepackage{xcolor}
\newcommand{\rev}[1]{{#1}}

\usepackage{url}

\journal{Image and Vision Computing}

\begin{document}

\begin{frontmatter}



\title{Unsupervised Domain Adaptation for Multitask Image Analysis in Realistic Context with Extreme Label Shift; Application to the CTAO first Large Sized Telescope} %% Article title


\author[a,b]{Michael Dell'aiera} %% Author name
\author[a]{Thomas Vuillaume} %% Author name
\author[b]{Alexandre Benoit} %% Author name

\affiliation[a]{organization={CNRS LAPP},%Department and Organization
            city={Annecy},
            postcode={74940}, 
            country={France}}

\affiliation[b]{organization={Université Savoie Mont Blanc LISTIC},%Department and Organization
            city={Annecy},
            postcode={74940}, 
            country={France}}

%% Abstract
\begin{abstract}
Unsupervised domain adaptation is a widespread set of methods that leverages the knowledge of a labeled source domain to train a model to perform well on a related unlabeled target domain. They generally introduce an auxiliary adaptation-related task that can be integrated into the multitask paradigm, which aims to merge multiple single-task models into a unified architecture. In this paper, we propose to associate domain adaptation and multitask balancing in the realistic context of an extreme class imbalance. Therefore, we propose a combined framework to cover and validate these approaches, and evaluate its performance in the physics-based context of the Cherenkov Telescope Array Observatory (CTAO). Along with a comparative study of some relevant adaptation techniques, we highlight the impact of extreme label shift and extend the investigations on importance weighting to rectify it. The complete code and results are published and available as open-source resources on Zenodo \cite{michael_dell_aiera_2024_13646001}.
\end{abstract}


% \begin{highlights}
% \item highlight1
% \item highlight2
% \end{highlights}

%% Keywords

\end{frontmatter}

% \linenumbers


\section{Introduction}
\label{sec:introduction}
Artificial intelligence has demonstrated its effectiveness during the last decades in various domains, and its performance has recently strengthened with the emergence of deep learning. Particularly in computer vision, convolutional neural networks \cite{lecun1998cnn} appeared to be a powerful tool for image analysis, although the quality of a model to converge depends on the reliability of the training dataset, as discussed in \cite{torralba2016unbiased}. In many real-world scenarios, and typically in physics applications, labeled acquisitions are either intrinsically not directly accessible or very hard to produce. Therefore, a pragmatic solution, classically adopted in particle physics, is to train a model on simulations, as it provides a controlled and labeled environment \cite{agostinelli2003, bernlohr2008simtelarray}. Nevertheless, simulations remain approximations such that the inference of the resulting model on the real-world observations generally leads to a decreased performance, as highlighted in \cite{jacquemont2021cta} regarding astrophysics applications. To tackle this problem, domain adaptation has been developed as a set of techniques to mitigate domain discrepancies \cite{zhao2020review}. However, their introduction to multitask balancing (MTB) becomes more complicated as each task is learnt at different rates \cite{kendall2018multitask}. Manually tuning the weights associated with each of these individual objectives is a costly process, and models may converge to a large variety of solutions depending on the selected hyperparameters. Emancipating from this labour is an arduous work but a first step consists in using some prior knowledge on the desired evolution of the weighting coefficients \cite{ganin2016domainadversarial}, for example gradually incorporating the contribution of the domain adaptation task over the iterations. 

Unsupervised domain adaptation (UDA) allows integrating unlabeled real-world data, referred to as the target, into the training procedure alongside with labeled simulations, referred to as the source. Nevertheless, class imbalance is a rather natural behaviour for many image analysis tasks, and can be particularly severe in fields like particle physics, where researchers are looking for rare events. In our work, we address UDA in a realistic scenario where the source is balanced but the target suffers from an extreme class imbalance. To our knowledge, this specific context has not been explored in the literature. In particular, we present the application of physics-guided deep learning to gamma-ray astronomy, which focuses on the detection and study of very high-energy gamma rays, the most energetic form of electromagnetic radiation. Specifically, we explore its integration in the Cherenkov Telescope Array Observatory (CTAO) first LST image analysis, which relies on Imaging Atmospheric Cherenkov Techniques (IACT) to detect gamma rays through the observation of Cherenkov light produced by the particle shower mechanism. As a result, the emitted photons are captured by the telescope camera optical system which generates a sequence of 40 snapshots. Usually, models input images of integrated charges and arrival times of the signal peak within the sequence.

In this paper, real data are replaced with controlled and degraded simulations to highlight the contribution of the two main discrepancies between the simulations and the telescope observations in the context of UDA, which are the label shift and the covariate shift. The former corresponds to the change of class occurrence between both domains. The latter, often approximated by the Night Sky Background (NSB) in the astrophysics community, or light pollution, contains the influence of external and atmospheric factors, such as stars or moonlight. Recent investigations have highlighted the significant role of NSB as one of the main contributors to domain shifts \cite{parsons2022investigations}. Moreover, training simulations are usually well balanced, as the particle classifier usually performs better when trained on a balanced dataset. On the other hand, real data typically have a ratio of at least $10^{-4}$ in favour of protons, and it changes not only with the observation coordinates of the sky, but also as a function of the particle energy, as precisely described in \cite{abe2023crab}. This difference in the label ratio is intrinsically a limiting factor of UDA \cite{liu2021adversarial}. \rev{These controlled degradations of simulated data are not intended to reproduce all discrepancies observed in telescope data. Rather, they provide a reproducible framework for isolating the dominant sources of mismatch identified in previous studies, allowing quantitative evaluation while preserving access to ground-truth labels.}

Relying on the multimodal and multitask deep learning model proposed in \cite{jacquemont2021cta}, the contributions of this paper are the following:
\begin{itemize}
    \item We compare the performance of multitasking in the context of UDA on both the classical digit images dataset and the LST simulation dataset.
    \item We introduce Conditional UDA (CDANN, CDeepJDOT, CDeepCORAL), an extension of the work of \cite{lipton2018detecting} to the extreme label shift inherent to gamma-ray astronomy. Our approach modifies the training procedure of the UDA task and significantly improves the model performance in the proposed challenging context.
\end{itemize}

This paper is organized as follows: UDA and MTB are first introduced in Section \ref{sec:related_work}. The proposed methods and the conditional UDA are presented in Section \ref{sec:method}. We first apply our methods to a standard UDA benchmark in Section \ref{sec:digits}. Then, we shed light on the importance of UDA for the CTAO image analysis in a controlled environment using simulations of the first LST. The summary of the obtained results are presented in Section \ref{sec:lst_simulations}. Finally, Section \ref{sec:conclusion} draws the conclusions, and paves the way to new perspectives in the application to real observations of gamma-ray sources.

In an effort for openness and reproducibility of our work and results, all the necessary code and results are collected and published in \cite{michael_dell_aiera_2024_13646001}.

\section{Related work}
\label{sec:related_work}

\subsection{Dataset shift}
Traditional machine learning relies on the assumption that test and training data are drawn from the same distribution. However, in practice, this ideal condition is rarely met, and biases occur. In its simplest definition, a dataset shift materializes when the training and test joint distributions, respectively denoted as $P_S(x,y)$ and $P_T(x,y)$, are different $\forall (x,y) \in \mathcal{X} \times \mathcal{Y}$. In our case, the image spaces are identical ($\mathcal{X}_S = \mathcal{X}_T = \Re^{d}$, with $d$ referring to the size of the data), so are the label collections $\mathcal{Y}_S$ and $\mathcal{Y}_T$. Depending on the origin of the discrepancy, it can be divided into multiple categories, as explained in \cite{kull2014patterns}.

\paragraph{\textbf{Covariate shift}}
Covariate shift occurs when the distribution of the input data $x$ changes, but the conditional distribution of the labels $y$ given the input $x$ remains the same.

\begin{equation}
    \begin{array}{rcl}
        P_S(x) & \neq & P_T(x) \\
        P_S(y|x) & = & P_T(y|x)
    \end{array}
\end{equation}

In IACT applications, it is mostly dominated by NSB variations \cite{parsons2022investigations}.

\paragraph{\textbf{Concept shift}}
Concept shift occurs when the conditional distribution of the labels $y$ given the input $x$ changes between the distributions. 

\begin{equation}
    \begin{array}{rcl}
        P_S(x) & = & P_T(x) \\
        P_S(y|x) & \neq & P_T(y|x)
    \end{array}
\end{equation}

\rev{Concept shift occurs when the conditional distribution of the labels $y$ given the input $x$ changes between the distributions. In the context of IACT, this may arise from imperfect simulation of atmospheric conditions, detector response, or calibration effects. However, these effects can be measured and mitigated with real observations using calibration and data-quality validation procedures, for example, by comparing distributions of low-level image parameters between simulations and observations.}

\paragraph{\textbf{Label shift}}
In general, there are no guarantees that the training source (simulations) and training target (real data) label distributions, respectively $p_S(y)$ and $p_T(y)$, are similar for all labels $y \in \mathcal{Y}$, and this case is referred to as label shift.

\begin{equation}
    \forall y \in \mathcal{Y}, P_S(y) \neq P_T(y)
\end{equation}

Most of the approaches of the literature rely on importance weighting, which consists in estimating the ratio of the source and target classes $\omega(y)$, and re-weight the target distribution accordingly \cite{liu2021adversarial, zhang2013domain, azizzadenesheli2019regularized, lipton2018detecting}, so that:

\begin{equation}
    P_S(y) = \omega(y) P_T(y)
\end{equation}

Although this approach yields increasing performance, it is only applicable when such ratio is still acceptable, that is to say each mini-batch contain enough statistics to estimate the weights $\omega$. In IACT, the particle flux follows different power laws and decreases exponentially as the energies grow. In addition, simulated training data must be balanced and representative for each class, making the field strongly affected by label shift.

\subsection{Domain adaptation}
Dataset shifts necessitate developing new methods to generalize well on related yet different domains. From a mathematical point of view, authors of \cite{wang2018survey} defines domain $\mathcal{D}$ as the combination of a feature space $\mathcal{X}$ and a marginal probability distribution $P(x)$ over a dataset $D$. As part of the transfer learning paradigm, domain adaptation consists in transferring knowledge from a labeled source domain to another distinct but related target domain. Depending on the availability of the target labels, it can be categorized as fully-supervised, semi-supervised or unsupervised, as proposed in \cite{zhao2020review}. In principle, it requires to invoke the covariate shift assumption stipulating that the conditional label distribution does not change between both domains, as explained in \cite{bendavid2010impossibility}. In the case of gamma-ray source detection, the parametric model $f^{\theta}$ allows retrieving the physical properties of the incident events, as $f^{\theta} = P_S(y|x) = P_T(y|x)$. 

\rev{Recent advances \citep{kim2022distilling, ngo2023} in Semi-Supervised Domain Adaptation (SSDA) have demonstrated the benefit of leveraging a small amount of labeled target data to improve domain alignment. These approaches achieve strong performance when a subset of labeled target samples is available, which is not possible in the context of IACT images. Moreover, covariate and label shift are expected to be the dominant shifts in this context.}

\rev{Consequently, we focus in this work on domain-alignment approaches that can be divided into four groups:}

\paragraph{\textbf{Discrepancy-based}} Firstly, discrepancy-based methods aim to minimize a measure of disparity between distributions. Although the computation of such distance or divergence theoretically requires the knowledge of the underlying distributions, neural networks offer the flexibility to estimate them stochastically. Various standard metrics exist, involving for example the computation of moments. In the case of the first-order statistics, it is referred to as Maximum mean discrepancy (MMD) \cite{tzeng2014mmd}, and its empirical approximation can be computed on the mean feature maps of the neural networks. \cite{long2015dan} introduces characteristic kernels to calculate the mean embedding of the feature vectors, highlighting that the application of multiple kernels contributes to an enhancement of the classification performance. Deep Correlation Alignment (DeepCORAL) \cite{sun2016deepcoral} aims to minimize the distance of the second-order statistics of the source and target latent features of size $d$, the correlation alignment quantifying the misalignment between the covariance matrices $C_S$ and $C_T$ through the minimization of the following objective function:

\begin{equation}
    \mathcal{L}_{DeepCORAL} = \dfrac{1}{4d^2} ||C_S - C_T||^2_{Frobenius}
    \label{eq:loss_deepcoral}
\end{equation}

As reducing the first- and second-order moments are not adapted for non-Gaussian distributions, Higher-order Moment Matching (HoMM) \cite{chen2019homm} unifies the framework by extending these approaches to higher-level statistics coupled with characteristic kernels. Similarly, Deep Joint Distribution Optimal Transport (DeepJDOT) \cite{courty2018deepjdot} relies on optimal transport to estimate the Wasserstein distance between both domains on the deepest latent features which, when minimized, ensures the overlapping of both distributions. The corresponding objective function is usually defined as:

\begin{equation}
        \mathcal{L}_{DeepJDOT} = \sum_{i,j} \pi_{i,j} C_{i,j}
        \label{eq:loss_deepjdot}
\end{equation}

where $\pi$ corresponds to the optimal transport plan and $C$ to the cost matrix. Domain-Transformer (DoT) \cite{chuan-xian2022dot} is an optimal transport related Transformer model that focuses on learning semantic consistency across both the source and the target domains to improve generalization of the classifier. Cross-Domain Transformers (CDTrans) \cite{xu2021cdtrans} relies on pseudo-labels and uses cross-attention to reduce the impact of label noise when labels are not correctly assigned to the target samples.

\paragraph{\textbf{Adversarial discriminative}} Secondly, adversarial discriminative models implement a domain discriminator along with a dual min-max objective function to learn domain-invariant features, such as Domain Adversarial Neural Network (DANN) \cite{ganin2016domainadversarial}. The feature extractor $G$ is complemented by an additional objective that is to mislead the domain classifier $D$, and therefore to ensure domain invariance while providing good performances on the other classification or regression task. As an adversarial optimization problem, it can be formulated using a Gradient Reversal Layer (GRL) $\mathcal{R}$, a pseudo-function that reverses the sign of the gradient during the backward pass. In practice, the propagated gradient returned from the domain classifier is weighted by an epoch-dependent function, ensuring the training of the feature extractor on the supervised task in the first place before incorporating the adaptation process smoothly over time. DANN objective function can thus be defined, using the Cross-Entropy (CE) metric, as:

\begin{equation}
    \mathcal{L}_{DANN} = \sum_{i} CE\left(D(\mathcal{R}\left[G(x_i)\right]), d_i\right)
    \label{eq:loss_dann}
\end{equation}

Furthermore, the domain classifier can be replaced with a domain metric. As the Wasserstein distance indeed has better properties regarding the gradients during backpropagation, it is considered as a promising tool to update and improve the model's performances \cite{shen2018wasserstein}. Such approach necessitates to impose a Lipschitz constraint on the domain metric, which can be easily implemented with a penalization on the derivative of the metric gradients \cite{gulrajani2017improved}. 

\paragraph{\textbf{Adversarial generative}} Oppositely, adversarial generative methods are based on Generative Adversarial Networks (GANs) \cite{goodfellow2014gan}, and aim to proceed image-to-image translation \cite{isola2016pix2pix}. COGAN \cite{liu2016coupled} is one type of such algorithm combining unpaired images with the use of two GANs synthesizing images from each domain. CycleGAN \cite{zhu2020unpaired} introduces the cycle-consistency constraint to perform unpaired image-to-image translation and ensure that the mappings are bijective. While discriminative models aims to obtain domain-invariant features, these approaches aim to project one distribution into another. Based on transformers, Image-to-image Translation with Transformers \cite{zheng2022ittr} is an unpaired image-to-image translation network following an encoder-decoder architecture and combines depth-wise convolutions along with self-attention into a hybrid perception block to highlight short- and long-range dependencies.

\paragraph{\textbf{Self-supervised}} Lastly, self-supervised methods integrate auxiliary self-supervised learning tasks into the primary task network, as co-training proves beneficial in reducing the dissimilarity between the source and target domains. Such methods, for example Multitask Auto-Encoder \cite{ghifary2015mtae}, learns to transform the initial distribution into analogues, so that the features that are extracted from the encoder are robust to domain variations. In a second place, these invariant features are used as the input of a classifier or regressor.

\subsection{Multitask balancing}
MTB refers to the simultaneous optimization of distinct but generally related tasks during the training step \cite{caruana2004multitask}. Analogously, learning a new language is easier when confronting speaking, writing and listening together. The classical approach for determining the loss weights that are in competition during the training process is conducted through a grid search optimization, and the choice of this hyper-parametrization has a critical impact on the performance of the model, even on a low-complexity dataset. The computational complexity thus increases exponentially as the number of tasks grows, shedding light on the need for task auto-balancing.

In addition to reducing computation costs, the advantage of multitask learning is twofold. Not only designing an architecture performing on multiple and complementary problems at the same time allows each task to benefit from each other \cite{he2017maskrcnn}, but also auxiliary tasks can be included to help constrain the problem and improve the overall performance \cite{jacquemont2021cta}. The global loss function thus becomes a weighted sum (using coefficients $w_t$) of the $T$ mono-objectives:

\begin{equation}
    \mathcal{L} = \dfrac{1}{T} \sum_{t=1}^{T} w_t \mathcal{L}_t
\end{equation}

Many automatic loss balancing methods have been proposed in the literature \cite{zhang2017survey_mtl}. Usually defined as baselines, Equal Weighting (EW) and Random Loss Weighting (RWL) \cite{lin2022rwl} are the most straightforward balancing procedure and respectively consist in applying the same coefficient to each task, or to sample them from a distribution, typically Normal. When the weights are tuned using a grid search procedure or with the expert's knowledge, they fall into the scope of Manual Weighting (MW).

The contribution of a task to the optimization of the model parameters can be quantified by the norm of its gradients. Depending on their implementations, certain tasks may have strong gradients at the beginning of the training, thereby driving parameters into a global minimum where the other tasks may potentially not optimize properly. In order to introduce task balancing, \cite{chen2018gradnorm} instantiates the GradNorm (GN) algorithm. This method aims to weight the task by a measure of their contribution through the determination of the inverse training rate $r_t$ and its projection on a common scale $\overline{G}_W(i) = \mathbb{E}_{\text{task}} \left(G_W^{(t)}(i)\right)$ by multiplying it with the norm of the gradients $G_W^{(t)}(i) = ||\nabla_W w_t(i) L_t(i)||_2$ calculated at the last shared layer $W$ at the iteration $i$. It corresponds to finding the weights $w_t$ that minimize the following GradNorm objective function:
\begin{equation}
    \mathcal{L}_{grad}(w_1, ..., w_T) = \sum_{t=1}^{T} \left| G_W^{(t)}(i) - \overline{G}_W(i) \left[r_t(i)\right]^{\alpha} \right|_{1}
\end{equation}
where $\alpha$ remains the last hyperparameter that tunes the power of the multitasking effect. For example, a value of $\alpha = 0$ indicates that the backpropagated gradients are forced to be equal at the layer $W$. The inverse training rate is $r_t$ defined as:
\begin{equation}
    r_t(i) = \dfrac{L_t(i)}{L_t(0)} \div \mathbb{E}_{\text{task}} \left(\dfrac{L_t(i)}{L_t(0)}\right)
\end{equation}
High values of $r_t$ signify that higher gradient norms are required in order to train the task more quickly.

Although GN only considers the norm of the gradients, their orientations appear to play a role in the quality of training. In the very high dimensional landscape of the weights, tasks can be in opposition if the gradients are not co-linear, thus their association can be counterproductive. A simple measure of conflict is the cosine similarity. \cite{yu2020pcgrad} proposes to solve conflicting gradients by projecting them into non-conflicting basis. \cite{guangyuan2022recon} identifies the shared layers that are conflicting the most frequently to change them to task-specific parameters.

Another relevant task balancing approach is Uncertainty Weighting (UW) \cite{kendall2018multitask} that relies on the task-dependent homoscedastic uncertainty to estimate the loss coefficients. This uncertainty can be defined as the standard deviation of the error distribution under the assumption that the associated likelihood follows either a Normal, Laplacian or Softmax distribution, which intrinsically limits the use case of this approach to a small selection of loss functions. The global objective loss can be derived as:
\begin{equation}
    \mathcal{L}(s_1, ..., s_T) = \sum_{t=1}^{T} K e^{-s_t} \mathcal{L}_t + s_t
\end{equation}
where $s_t = log(\sigma_t^2)$ defines the log-variance. The constant $K$ depends on the likelihood function, with $K = 2$ for Normal or Laplacian distributions and $K = 0.5$ for the SoftMax distribution.

\subsection{Multitask learning domain adaptation}
There are many applications of UDA combined with MTB in the literature \cite{ren2018umtda, yang2021umtda, zhang2019umtda}, but these approaches all rely on the same strategy, which is to manually balance the tasks by optimizing the coefficients using MW. To the best of our knowledge, there is no work trying to associate UDA and MTB using GN or UW.

\section{Method}
\label{sec:method}

\begin{figure*}[ht!]
    \centering
    \includegraphics[width=\linewidth]{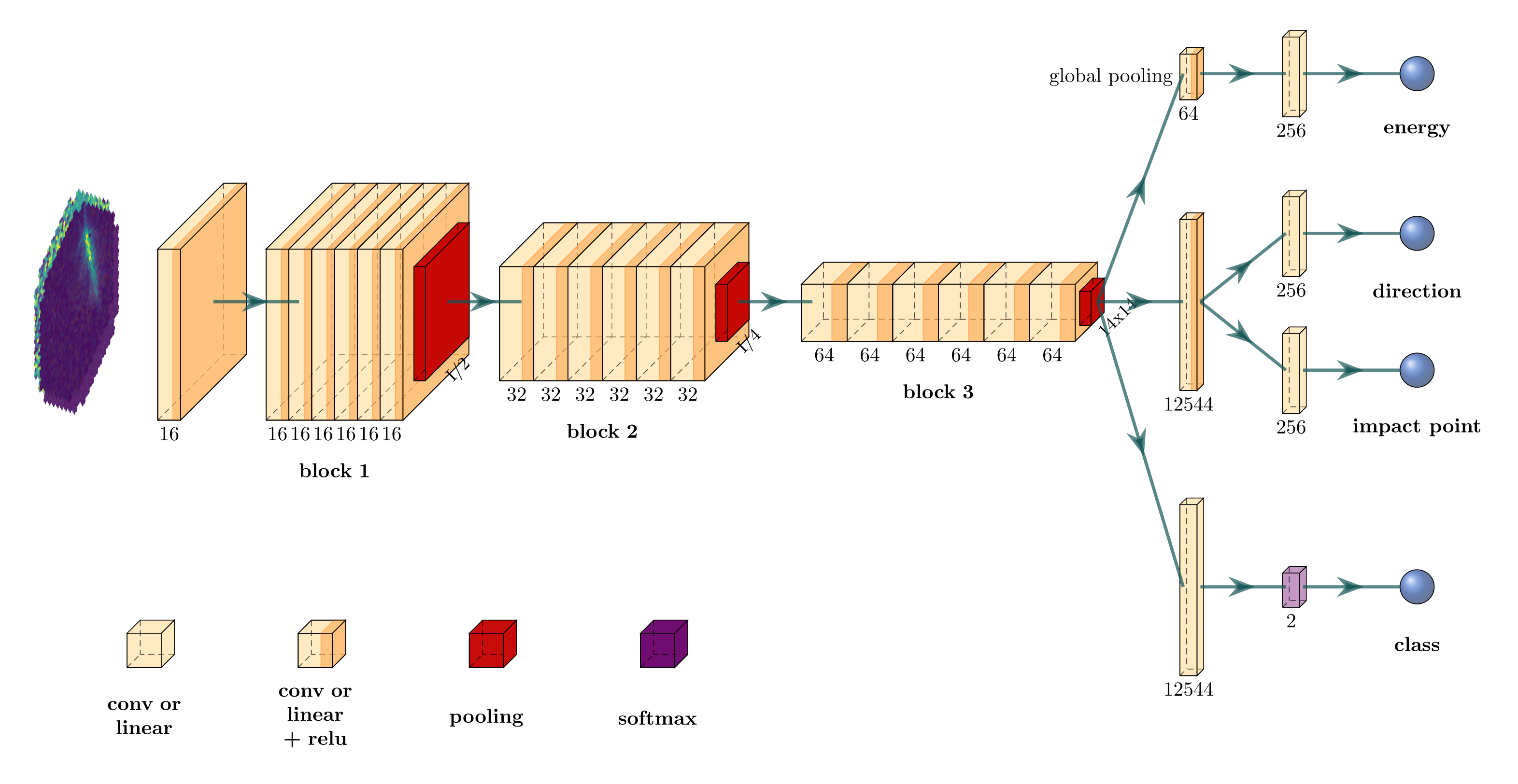}
    \caption{The $\gamma$-PhysNet architecture. This illustration represents the feature maps that output from the convolution filters or the fully-connected layers (that can be associated with a non-linearity) of the network. For example, \textit{conv} designates the feature maps that output a convolution layer. The backbone is composed of three blocks and the identity mapping corresponds to full-activation \cite{he2016identity}.}
    \label{fig:architecture}
\end{figure*}

\subsection{Motivation}
We propose in the following to apply our methodology to gamma-ray astronomy with IACT.  The detection workflow of gamma-ray sources for the first LST is currently based on a conventional machine learning approach. It relies on feature extractions of the signal contained in the integrated images. This is performed with the Hillas algorithm \cite{hillas1985hillas}, while the reconstruction of the shower-initiating particle parameters are performed subsequently using Random Forest \cite{ohm2009bdt}. This procedure, referred to Hillas+RF, operates under the assumption that the footprint of the integrated signal can be approximated with an ellipsoidal shape. It can be characterized with a few simple parameters, among which the moments - up to the third order - of the light distribution on the camera. 

Thanks to a cleaning procedure that removes the background noise from the images, Hillas+RF exhibits great robustness properties and results on the detection of real sources \cite{abe2023crab}. However, the cleaning procedure is imperfect, and the model performance degrades at lower particle energies or in the presence of strong NSB, as it is difficult to extract relevant morphological information from a few pixels, or if the signal-to-noise ratio is not favourable. Nevertheless, detecting gamma rays at lower energies is crucial as they carry the majority of the gamma-ray flux \cite{abe2023crab}. Performing poorly on this portion of the electromagnetic spectrum ultimately decreases the detection capabilities, limiting the quantity of data available for the analysis. In such scenarios, the use of deep neural networks can potentially greatly improve the performances at lower energies. Moreover, since deep learning is not bound to the Hillas features, it is capable in principle to learn better representations at intermediate layers.

The $\gamma$-PhysNet is the first full-event reconstruction neural network of the CTAO for the detection of gamma-ray sources \cite{jacquemont2021cta}. As illustrated in Figure \ref{fig:architecture}, it is composed of two entities that are respectively a ResNet \cite{he2015resnet} feature extractor, augmented with attention mechanisms \cite{hu2019squeezeandexcitation}, and a multitask architecture for the reconstruction of an incident particle physical parameters (namely its class, energy, arrival direction and impact point on the ground). Because it is by essence impossible to obtain labeled telescope acquisitions, the training of the network relies on simulations. However, real observations are affected by environmental conditions, such as the presence of the moon, clouds or Calima. Moreover, variations in the position in the sky of the gamma-ray source result in different atmospheric thicknesses traversed by the particle shower, thereby impacting the physical attribute reconstruction. Consequently, the proximity of the observed direction to the coordinates of the training dataset also plays a significant role. 

In general, neural networks are sensitive to these phenomena, and it translates, in our case, into biases in the reconstruction of the position of the source, or into degraded performance \cite{jacquemont2021cta}. The current solution consists in modifying the training data based on the observations by matching the background levels, but this run-wise procedure suffers by construction from a low generalization. Hence, the search for a more universally applicable method becomes imperative, and new approaches have been proposed using UDA as a first step in that direction \cite{dellaiera2023uda}.

\subsubsection{Method selection}
In comparison to the current methodology of CTAO data analysis, the originality of the UDA is the inclusion of real observations into the training procedure along with the simulations. In this work, we selected three UDA approaches that are DANN \cite{ganin2016domainadversarial}, DeepJDOT \cite{courty2018deepjdot} and DeepCORAL \cite{sun2016deepcoral}. These selected methods fall into the adversarial discriminative and discrepancy-based frameworks. Firstly, DANN is one of the most popular approaches for UDA, has been successfully applied in a wide variety of contexts, and is supported by a strong mathematical background. Secondly, the Wasserstein distance exhibits excellent convergence properties compared to other distance between distributions, as demonstrated in  \cite{arjovsky2017wasserstein}, such as Jensen-Shannon or Kullback-Leibler divergence; thus DeepJDOT is a relevant choice to reduce domain discrepancy. Lastly, as the NSB is known to be one of the main contributors to the discrepancies between simulations and real data, minimizing the difference between the first- and second-order statistics seems to be a relevant strategy, making DeepCORAL an interesting choice for our application. In particular, the NSB is usually approximated with a Poisson noise, which rate can be estimated with the mean or variance over the samples. Overall, these three methods constitute a relevant baseline to start building more sophisticated algorithms. Adversarial generative methods face limitations because of the unequal ratio of particles in the real acquisition dataset, which precludes cycle-consistency. Self-supervised-based methods, through transformer models \cite{he2021mae}, are currently being investigated for a future work and are not presented here. 

\subsubsection{Merging domain adaptation and multitask balancing}
\begin{figure}
    \centering
    \includegraphics[width=\columnwidth]{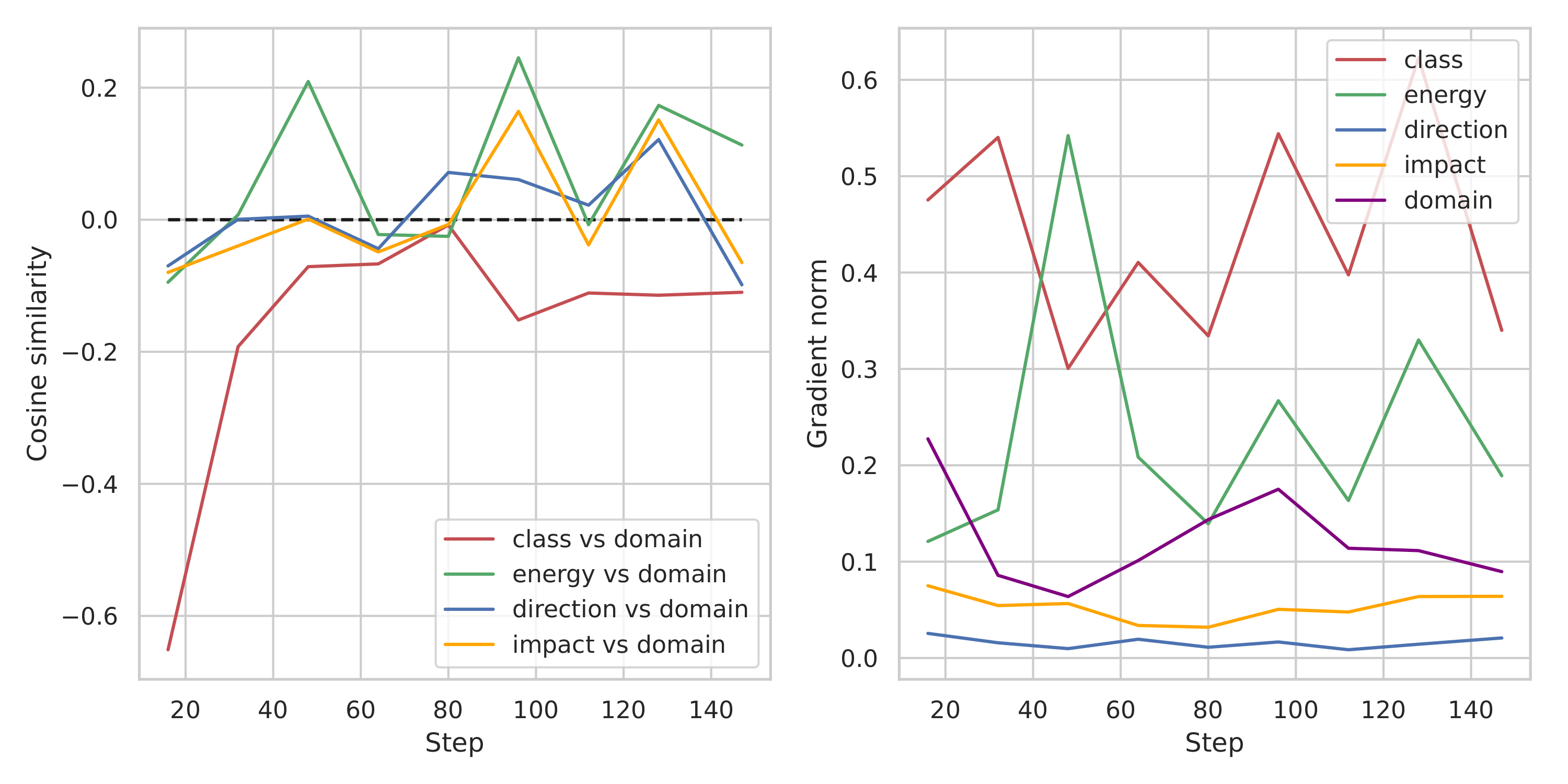}
    \caption{The cosine similarity between the domain task and the full-event reconstruction tasks is shown on the left panel. The gradient norms of each task are plotted on the right panel. In both cases, the domain adaptation method is DANN. Conflicting tasks, for example \textit{class} versus \textit{domain} (depicted in red) exhibit a negative similarity which, associated with strong gradients, may affect the quality of convergence of the model.}
    \label{fig:conflicting_task}
\end{figure}

Including UDA into the MTB framework aims at integrating the associated loss coefficients within the automatic determination of the weights of each task. However, the analysis of the amplitude and cosine similarity of the gradients at the last shared layer highlights conflicting gradients between the UDA task and the reconstruction of the physical attributes, as depicted in Figure \ref{fig:conflicting_task}, turning the optimization into a difficult process. Strong gradient amplitudes from the domain task can be detrimental for the performance on the source dataset when tasks are in opposition. Inspired by DANN, the strategy of progressively including the contribution of UDA to the training with the use of a gradient weighting procedure can be extended to DeepJDOT and DeepCORAL through the definition of a gradient layer (GL) $\mathcal{G}$, which aims at weighting the domain adaptation gradient during the backward pass. It can be reversed in the case of DANN.
\begin{equation}
    \mathcal{G}(x) = x \ \ \text{ and } \ \ \dfrac{d\mathcal{G}}{dx} = (-1)^R f(i) \times I
    \label{eq:gradient_layer}
\end{equation}
where $R = 1$ for reversal else $R = 0$, $f$ is a function of the current iteration $i$ defined by the user, and $I$ is the identity matrix. Commonly, $f$ is chosen as the function that progressively evolves from $0$ to $1$, for example:

\begin{equation}
\label{eq:gradient_func}
    f(i) = \dfrac{2}{1 + \exp{\left(-\gamma g(i)\right)}}  - 1 \in [0, 1]
\end{equation}

where $g(i) = \dfrac{i}{\text{max epoch}} \in [0, 1]$ defines the temporal normalized evolution of the training procedure.

\subsubsection{The dataset shift challenges}
\begin{figure}
    \centering
    \includegraphics[width=\columnwidth]{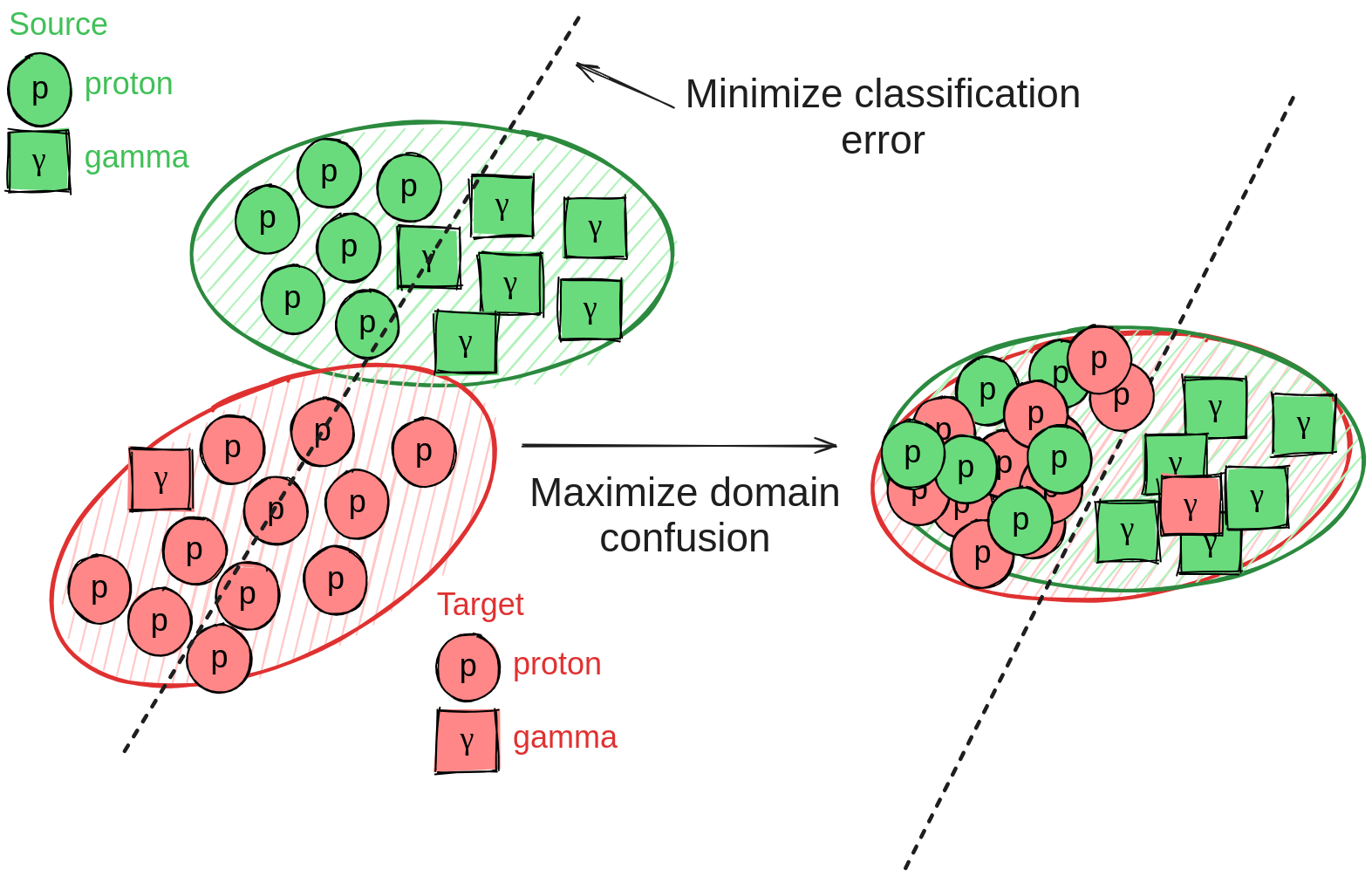}
    \caption{Illustration of UDA. While maintaining good performance on the source data with the minimization of the classification error, the addition of the domain confusion task allows to find a shared representation of both domains. This objective is more difficult when accompanied with label shift, that is to say the unbalanced ratio of particle classes across both domains, in our case protons and gammas. In real life scenarios, this ratio can reach $10^{4}$ in favor of protons.}
    \label{fig:domain_confusion}
\end{figure}

Figure \ref{fig:domain_confusion} illustrates the concepts of dataset shifts in the context of UDA. Covariate shift is dominated by the NSB differences, as it has been demonstrated to account for the main source of discrepancies between simulations and real observations \cite{parsons2022investigations}. Recently, UDA has been proposed in a previous contribution to tackle NSB in the context of LST simulations \cite{dellaiera2023uda}. Authors introduce target data that are perturbed with the addition of an artificial Poisson noise in order to mimic the real observations. They concluded that, although the proposed methods compensate for the introduced noise across almost all energy ranges, a gap in performance still persists on the energy resolution and bias at lower energy range of the calculated IRFs. Furthermore, these approaches do not consider the label shift.

Another challenging issue in transitioning from simulations to real data while applying UDA is the label shift, that is to say the difference of the class ratio between gamma and protons in the source and target datasets. Although they are equally represented in the training simulations, real data usually contains less than $1$ gamma for $10^{4}$ protons. Thus, for mini-batches of a typical size of $10^{2}$ samples, the probability of having no gamma in the target mini-batch is considerable. Thus, it becomes crucial to adapt the algorithms to take into account such label shift.

Following the work of \cite{liu2021adversarial} on importance weighting, we introduce the conditional versions of DANN (CDANN), DeepJDOT (CDeepJDOT) and DeepCORAL (CDeepCORAL) to tackle the strong label shift encountered gamma-ray astronomy. As the flux of gammas and protons can be evaluated from previous observations, it is possible to evaluate the value of the class-wise balancing parameter $\omega \in \mathbb{R}^2$. In fact, if $\omega = \left(\begin{array}{c}
    \omega(\gamma) \\
     \\
    \omega(p)
\end{array}\right)$, the weighting coefficient can be determined as:

\begin{equation}
    \begin{array}{ll}
        \omega(\gamma) & = \dfrac{P_T(\gamma)}{P_S(\gamma)} = \dfrac{n^{\gamma}_T}{n^{\gamma}_S} = \dfrac{\epsilon}{n^{\gamma}_S} \approx 0 \\
         & \\
        \omega(p) & = \dfrac{P_T(p)}{P_S(p)} = \dfrac{n^{p}_T}{n^{p}_S} = \dfrac{n^{p}_S + n^{\gamma}_S - \epsilon}{n^{p}_S} > 1
    \end{array}
\end{equation}

where $\epsilon = n^{\gamma}_T$ refers to the residual amount of gammas in the target set, and $n^{p}_T$, $n^{\gamma}_S$, $n^{p}_S$ refer to the number of gammas $\gamma$ or protons $p$ in the target or source set. Moreover, in the training process, the mini-batch are created to contain the same number of source and target images ($n_S = n_T$). Then, the composition of the mini-batch respects the following equation:

\begin{equation}
    n_S = n^{\gamma}_S + n^{p}_S = n^{\gamma}_T + n^{p}_T = n_T
\end{equation}

From a training point of view, this is equivalent to minimize the discrepancy between the simulated protons and the whole target dataset. 

In the following, we denote $D_S = \{p_S^{(i)}, \gamma_S^{(j)}\}_{i,j}$ the finite source samples containing the source protons $p_S^{(i)}$ and the source gammas $\gamma_S^{(j)}$. Similarly, the target finite samples are defined as $D_T = \{p_T^{(i)}, \gamma_T^{(j)}\}_{i,j} = \{x_T^{(k)}\}_k$, as they are undifferentiated by nature.

Specifically, for DANN, the domain classifier loss is masked based on the source particle labels, and the loss function becomes:

\begin{equation}
    \mathcal{L}_{CDANN} = \sum_{i \text{ }|\text{ } x_i \in \{p_S, x_T\}} CE\left(D(\mathcal{R}\left[G(x_i)\right]), d_i\right)
    \label{eq:loss_cdann}
\end{equation}

In the case of DeepJDOT, the cost matrix is computed only between protons in the source mini-batch and a sampled target mini-batch to maintain square arrays. The optimal transport plan is then deduced afterwards from these selected samples, following the two-step procedure proposed in \cite{courty2018deepjdot}. If $G$ defines the feature extractor of the neural network, then the cost function is described as:

\begin{equation}
        C_{i,j} = ||G(x_i) - G(x_j)||^2, (x_i, x_j) \in \{p_S^{(i)}\}_i, \times \{x_T^{(j)}\}_{j==i}
        \label{eq:loss_cdeepjdot}
\end{equation}

Finally, DeepCORAL computes statistics using source protons and the entire target mini-batch. The source covariance matrix is then computed as:

\begin{equation}
    C_S = \dfrac{1}{m - 1} \left(D_S^T D_S - \dfrac{1}{m}(\text{1}^T D_S)^T(\text{1}^T D_S)\right)
    \label{eq:loss_cdeepcoral}
\end{equation}

where $D_S = \{p_S^{(i)}\}_{i=1}^{m}$ refers to the mini-batch containing $m$ source protons. The calculation of the target covariance matrix is left unchanged. In practise, $m$ is mini-batch-dependent for both CDeepCORAL and CDeepJDOT, as the number of source protons changes at each iteration.

Compensating for the label shift can also be effectuated by integrating simulated gammas into the target dataset. Despite the fact that it ultimately introduces bias, this could be considered as promising because synthetic gammas accurately represent observed data; thus the discrepancy with real data mostly relies on the NSB. As mentioned in \cite{dellaiera2023uda}, this situation, referred to as $\text{MC}^{*} \rightarrow \text{MC}^{*}$, yields the same performance as $\text{MC} \rightarrow \text{MC}$. However, the gamma/proton ratio in a gamma-ray emitter observation is never known in advance, as it is a source-dependent property. In fact, some considered candidates might not produce any gamma radiation at all and a bias towards gammas in the target set must be prohibited to not produce any false detection. More practically, if the particle ratio is as low as $10^{-4}$, there is no expected impact of gamma particles on the domain confusion task.

\section{Benchmarks and validation on digits classification}
\label{sec:digits}
In this section, we first validate our approaches on the standard digit domain adaptation datasets and compare the different selected multitask strategies combined with our models. We then demonstrate the relevance of the GL in the case of DeepJDOT.

\subsection{Dataset}
The digit classification domain adaptation benchmark consists of four different data collections: MNIST \cite{lecun1998gradient}, USPS \cite{usps}, MNISTM \cite{ganin2016domainadversarial} and SVHN \cite{svhn}. These data collections are an interesting use case as they represent different levels of difficulty: the first two datasets are closely related and the digits lay on a black background, whereas the last two contain much more information with a complex RGB background. In addition, SVHN images can contain multiple digits, but only the one in the image centre must be classified, strengthening the complexity of the analysis. A more detailed description of the datasets is given in Table \ref{tab:digits_dataset}. Each dataset sample is potentially pre-processed to provide 3 colour channels with the replication of the unique channel, and images are adjusted to 28x28 pixels using either zero-padding or subsampling with bilinear interpolation. Overall, this case study solely illustrates the covariate shift problem, as the digit datasets contain no shift in the labels.

\begin{table}[h]
\centering
\begin{tabular}{ p{2.2cm}||p{2.0cm}|p{2.0cm}|p{2.0cm}|p{2.0cm} }
 & MNIST & MNISTM & SVHN & USPS \\
\hline
\hline
Resolution & 28x28x1 & 28x28x3 & 32x32x3 & 16x16x1 \\
\hline
Transforms & Duplicate channel & None & Resize to 28x28 & Zero-padding, duplicate channel \\ 
\hline
\#Training & 48000 & 48000 & 14651 & 5833 \\
\hline
\#Validation & 12000 & 12000 & 58606 & 1458 \\
\hline
\#Test & 10000 & 10000 & 26032 & 2007 \\
\end{tabular}
\vspace{0.2cm}
\caption{Meta-data of the digits datasets and their related preprocessing.}
\label{tab:digits_dataset}
\end{table}

\subsection{Hyper-parametrization}
We followed the ablation study proposed in \cite{courty2018deepjdot} to validate our approach. For all the considered experiments, the model is identical and consists in a feature extractor - a cascade of six $3 \times 3$ convolutional layers implemented with $(8, 8, 16, 16, 32, 32)$ filters, followed by an average pooling of size 5x5 - and a classifier - a fully connected layer with a softmax normalization. For DANN, the domain classifier is composed of two fully connected layers of size 100 and 2. The Adam optimizer with a learning rate of $1e^{-3}$ is used to update the models for 50 epochs. Batch size is set to 256 for both domains. In total, DeepJDOT, and DeepCORAL count 26306 parameters and DANN counts 106808 parameters. Batch Norm (BN) is used as normalization layers and the selected non-linearity is ReLU. The domain adaptation gradients are weighted using the GL with parameters $\gamma = 10$ when applied. The log-variances of UW and the parameters of GN are updated using the stochastic gradient descent (SGD) optimizer with no momentum. All optimizers have a learning rate of $1e^{-3}$ and a weight decay of $1e^{-4}$. A learning rate scheduler is employed and consists in reducing the learning rate by a factor of $10$ every 10 epochs.

Finally, each experiment is repeated with ten different seeds to account for the variability in parameter initialization and data shuffling. We thus report, in the results section, the average measures over these ten repetitions for each configuration.

\subsection{Results}
The classification accuracy obtained from the considered approaches are presented in Table \ref{tab:digits_results}. In order to highlight the contribution of MTB and domain gradient weighting, a unified training and evaluation procedure is required. This induces a slightly different implementation of the models and the selection of hyperparameters (number of training epochs and the learning rates), compared to the state-of-the-art results described by \cite{courty2018deepjdot}. Across all trials, MW is conducted using the weighting coefficients $\lambda_{class}$ and $\lambda_{domain}$ calculated with a grid search procedure based on MNIST$\rightarrow$MNISTM only.

Firstly, we compare MW with UW and GN for each of the selected methods. On the one hand, DANN and DeepCORAL combined with UW either reach or outperform MW, sometimes by a significant margin ($+6\%$ for DANN on USPS$\rightarrow$MNIST). On the other hand, DeepJDOT's best performance is obtained either with GN or UW depending on the scenario, but the gain in performance can be remarkable ($+10\%$ for DeepJDOT on SVHN$\rightarrow$MNIST). This simple example shows that it is possible to use MTB not only to save on computing time but also to improve the performance of the model. Moreover, it is important to note that, even though DeepJDOT and DeepCORAL cannot be mathematically integrated into the UW theory, as they cannot be considered as likelihood following a Normal or Laplacian distribution, they still perform well in this context of digits datasets.

Furthermore, for fair comparison, all the methods rely on the same hyperparametrization, but it is noticeable that the performance of UDA on the SVHN$\rightarrow$MNIST scenario is relatively low for both DANN and DeepJDOT, especially compared to the literature results. However, doubling the number of training epochs and removing the learning rate scheduler increase the performance of DANN by 20 points for MW, UW and GN, finally reaching the state-of-the-art performance. This case illustrates a slower convergence of the considered methods. For DeepJDOT, we did not manage to retrieve the results of \cite{courty2018deepjdot} in this particular scenario, but we noticed a high sensitivity to the training hyperparameters on the performances of this approach.

\renewcommand{\arraystretch}{1.5} % 1.5 times the default row spacing
\begin{table*}
\centering
\begin{tabular}{ p{2.4cm}|p{1cm}||p{1.9cm}|p{1.9cm}|p{1.9cm}|p{1.9cm} }
\hline
\multicolumn{6}{c}{Source $\rightarrow$ Target} \\
\hline
Method & MTB & MNIST $\rightarrow$ USPS & USPS $\rightarrow$ MNIST & SVHN $\rightarrow$ MNIST & MNIST $\rightarrow$ MNISTM \\
\hline
\hline
Vanilla & - & $0.89$ & $0.75$ & $0.54$ & $0.26$ \\
\hline
\hline
DANN & MW & $0.90$ & $0.86$ & $\textbf{0.55}$ & $0.95$ \\
\hline
DANN & UW & $\textbf{0.95}$ & $\textbf{0.92}$ & $\textbf{0.55}$ & $\textbf{0.97}$ \\
\hline
DANN & GN & $0.92$ & $0.89$ & $0.49$ & $0.96$ \\
\hline
\hline
DeepJDOT & MW & $\textbf{0.95}$ & $\textbf{0.94}$ & $0.65$ & $0.92$ \\
\hline
DeepJDOT & UW & $0.92$ & $0.92$ & $\textbf{0.75}$ & $\textbf{0.97}$ \\
\hline
DeepJDOT & GN & $\textbf{0.95}$ & $\textbf{0.94}$ & $0.69$ & $0.89$ \\
\hline
\hline
DeepCORAL & MW & $0.94$ & $\textbf{0.90}$ & $0.61$ & $0.79$ \\
\hline
DeepCORAL & UW & $\textbf{0.95}$ & $\textbf{0.90}$ & $0.61$ & $\textbf{0.82}$ \\
\hline
DeepCORAL & GN & $0.94$ & $0.89$ & $\textbf{0.62}$ & $0.72$ \\
\hline
\end{tabular}
\vspace{0.2cm}
\caption{Ablation study of DANN, DeepJDOT and DeepCORAL corresponding to the mean accuracy over ten seeds on the target dataset. The source accuracy is not mentioned. Only are reported the best performing strategies, but the impact of the GN $\alpha$ hyperparameter and of the gradient weighting are described in the next sub-section.}
\label{tab:digits_results}
\end{table*}

\subsubsection{Impact of GradNorm hyper-parameter $\alpha$}
Despite GN reduces the need for costly grid search optimization of the weights, the influence of its hyperparameter $\alpha$ must be explored. We evaluate the performance sensitivity of our selected UDA methods to $\alpha$ using a selection of $\alpha \in \{0.1, 0.5, 1.5, 3.0\}$. Experimentally, as presented in Table \ref{tab:digits_alpha}, we obtain that in the case of the digits datasets, our selected methods seem insensitive to the value of $\alpha$ on average. However, a finer analysis highlight differences depending the scenarios. DANN on MNIST$\rightarrow$USPS and USPS$\rightarrow$MNIST works best for low values of $\alpha$ ($\alpha = 0.1$), but is not influenced on MNISTM$\rightarrow$MNIST. DeepJDOT performance are at best on SVHN$\rightarrow$MNIST with $\alpha = 0.5$ on the source, but with $\alpha = 1.5$ on the target. DeepJDOT is less sensitive to $\alpha$ when the gradient weighting is used, while DANN is more sensitive to this augmentation. DeepCORAL appears less sensitive with or without GL.

\renewcommand{\arraystretch}{1.5} % 1.5 times the default row spacing
\begin{table}
\centering
\begin{tabular}{ c|cccc }
 & $0.1$ & $0.5$ & $1.5$ & $3.0$ \\
\hline
\hline
DANN  & \textbf{0.87} & 0.86 & 0.86 & 0.86 \\
\hline
DeepCORAL  & \textbf{0.86} & \textbf{0.86} & \textbf{0.86} & \textbf{0.86} \\
\hline
DeepJDOT  & 0.90 & \textbf{0.91} & \textbf{0.91} & \textbf{0.91} \\
\end{tabular}
\vspace{0.2cm}
\caption{Mean accuracy across all experiments while varying the GradNorm hyperparameter $\alpha$. Higher is better. In this table, DeepJDOT is combined with GL.} 
\label{tab:digits_alpha}
\end{table}

\subsubsection{Impact of the Gradient Layer}
\begin{figure}
    \centering
    \includegraphics[width=\columnwidth]{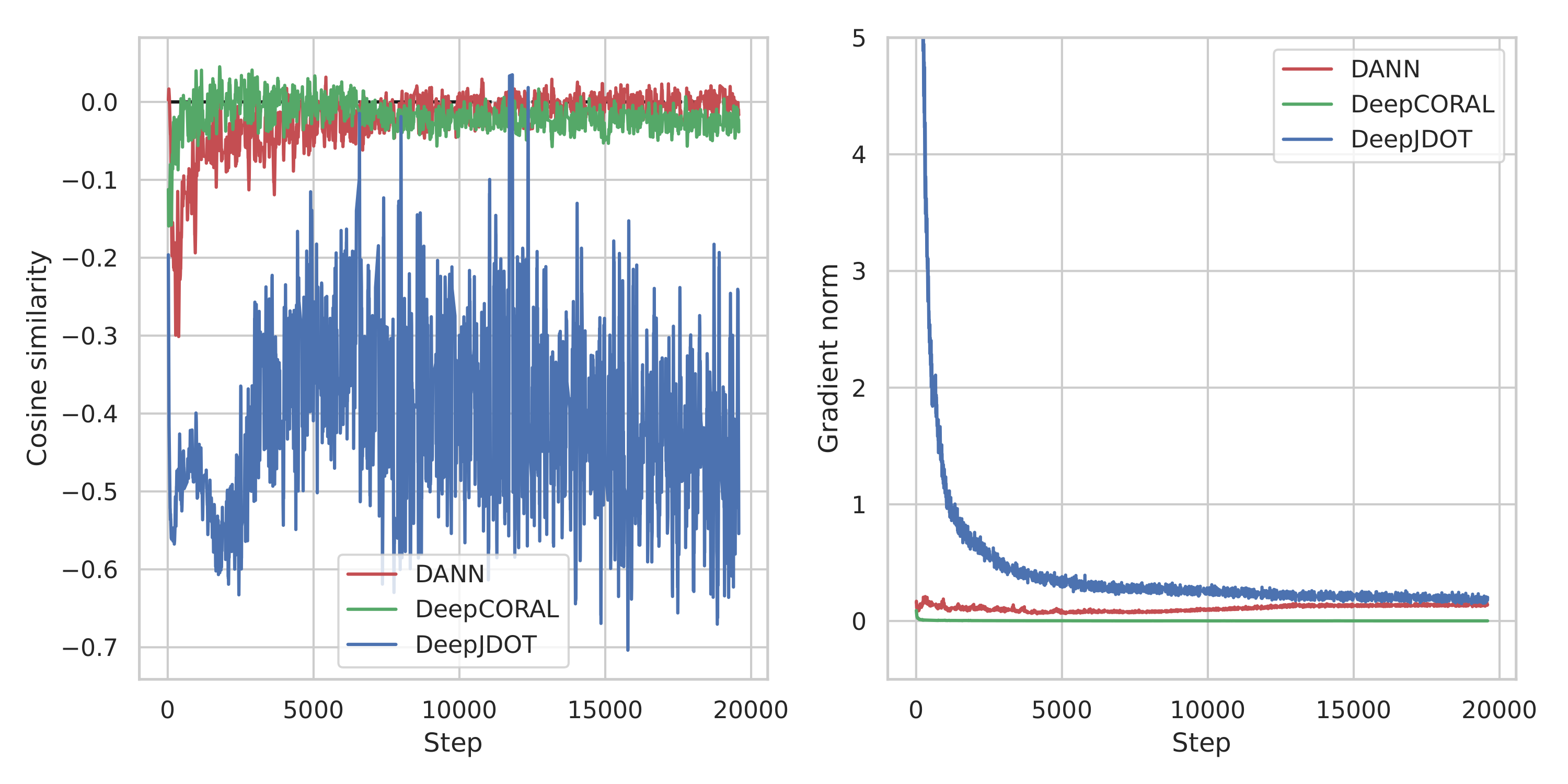}
    \caption{Cosine similarity between the domain confusion task and the digit classification task (left), and the gradient norms (right), using DANN, DeepCORAL and DeepJDOT on the SVHN$\rightarrow$MNIST scenario. In the case of DeepJDOT, the domain and digit classification tasks are highly conflicting. Moreover, the norm of the DeepJDOT gradients is very high at the beginning of the training, suggesting that the domain confusion task is not optimize properly.}
    \label{fig:conflicting_task_digits}
\end{figure}

The gradient layer consists in integrating an additional pseudo-function (see Eq. \ref{eq:gradient_func}) to the domain adaptation branch to better control the resulting domain adaptation gradient during the backpass. As illustrated on Figure \ref{fig:conflicting_task_digits}, the Wasserstein distance of DeepJDOT provides strong gradient values at the beginning of the training, which may lead the model to optimize very fast on the domain adaptation task to the detriment of the other tasks. 

Weighting the domain adaptation gradients at the beginning of the training allows the model to optimize on the source data first, which appears to be the best strategy for DeepJDOT in this specific context. On the contrary, it reduces the performance of DANN and DeepCORAL, which work better without gradient weighting. A possible reason is that our model uses a learning rate scheduler, which reduces the contributions of the weight update over time, which can be detrimental when combined to GL in the case of DANN and DeepCORAL.

\renewcommand{\arraystretch}{1.5} % 1.5 times the default row spacing
\begin{table}
\centering
\begin{tabular}{ c|cccccc }
  & UW & $0.1$ & $0.5$ & $1.5$ & $3.0$ \\
\hline
\hline
DANN & 0.05 & -1.56 & -1.56 & -0.83 & -2.48 \\
\hline
DeepCORAL & -0.47 & -0.81 & -0.73 & -0.88 & -0.91 \\
\hline
DeepJDOT & 21.8 & 18.9 & 17.1 & 15.1 & 15.0 \\
\end{tabular}
\vspace{0.2cm}
\caption{Performance gap between the base version and the GL-augmented version. The results correspond to the difference of both mean accuracies across all experiments. Values $\sim 0$ signify that both versions perform similarly on average. Values $>0$ highlight a globally beneficial results of GL, while values $<0$ indicate that no layer yields better performance.}
\label{tab:digits_gl}
\end{table}

\subsubsection{Conclusion on the digit ablation study}
The performance of our models on the digit datasets evinced multiple conclusions. Firstly, incorporating MTB with UDA allows us to retrieve or even outperform the results obtained with MW, without the need for costly grid search procedures. Secondly, it highlights the gradient conflicts between both the digit and domain classification tasks. Nevertheless, this issue can be partially resolved with the use of a GL. Currently, these affirmations hold solely in the specific case of the digit datasets, and the next section aims to generalize our results in a more complicated particle physics scenario, with an additional complexity introduced by extreme label shift.

\section{Application to the CTAO first LST simulations}
\label{sec:lst_simulations}
In this section, we benchmark the considered UDA approaches on our astrophysical problem related to gamma-ray analysis. The aim is to estimate the physical parameters of gamma ray incident particles, namely the energy, the orientation and the particle type (gamma or proton). The shower axis reconstruction, although not a part of the full-event reconstruction, is an auxiliary task that helps constrain the problem and improves the performance of the model \cite{jacquemont2021cta}. Compared to the preliminary experiments presented in the previous section, this case study mixes both classification and regression tasks and, importantly, the label shift problem.

\subsection{Datasets}
We consider the LST project dataset that simulates gamma and proton events as observed by the first LST. It relies on Monte-Carlo simulations generated with CORSIKA \cite{heck1998corsika} and \texttt{sim\_telarray} \cite{bernlohr2008simtelarray} at a pointing zenith angle of $20^{\circ}$ and azimuth $180^{\circ}$, pre-processed with cta-lstchain \cite{lopez2022lstchain} using pixel-wise charge integration as described in \cite{abe2023crab}. The dataset is here referred to as Prod5 LST-1 mono-trigger. This results, for each event, in pairs of inputs corresponding to the integrated charge amplitude and the photons average arrival time for each pixel. Such bi-modal samples are used as input for our network. As the images lie at the sensor level on a non-conventional hexagonal grid of pixels, we interpolate them on a regular grid using the dl1-data-handler library \cite{kim2022dl1dh}. We thus benefit from the high-performing implementations of the PyTorch convolution, reducing the training computational time, while obtaining the same performance compared to a more sensor tailored approach that relies on indexed convolutions \cite{jacquemont2019indexed} for hexagonal pixel images.

Complying with previous works on CTAO event reconstruction \cite{dellaiera2023uda, jacquemont2021cta}, the training and validation splits respectively correspond to $80\%$ and $20\%$ of the training set. The training set contains 1.1M diffused gammas and 0.76M protons whereas the test set contains 1.1M point-source gammas mimicking a gamma-ray point source and 0.76M protons.

As a validation of our approach and in order to fully constrain and control the problem, we create the target dataset from the LST simulations. It represents more realistic image acquisitions, taking advantage of recent knowledge from real sensor behaviours and the data distribution shift with the standard simulations. In more detail, it is performed with the addition of a random noise in the images following a Poisson distribution of rate of $\delta \lambda = 0.46$, as conducted in the analysis of \cite{jacquemont2021cta}. This value highly depends on the considered observations, and is chosen in this paper in accordance to previous work \cite{dellaiera2023uda}. Moreover, even though the target labels are not used during training, we take advantage of such controlled simulations in the evaluation procedure to assess each method performance using the standard figures of merit presented in the next section.

\subsection{Figures of merit}
Evaluating the performance of a model in the context of IACTs is possible using Instrument Response Functions (IRFs) that are computed using the test labeled simulations as a function of the true particle energies. The expected performances for the complete array of CTAO are presented on the CTAO webpage\footnote{\url{https://www.cta-observatory.org/science/ctao-performance/}} \cite{ctao2021fom}. This work utilizes a subset of those metrics relevant to our study that are described in this section and used to compare the methods presented here. As such, they are not representative of the absolute performances of CTAO as the datasets and event selection might differ from other works. The IRFs calculation is similar to \cite{abe2023crab}.

\paragraph{Gamma classification}
The gamma/proton classifier outputs a single digit for each event, the gammaness, that can be seen as the probability for that event to be a gamma event. The gammaness is then used to compute the Receiver Operating Characteristic (ROC) function. Here, we present as a integrated metric the Area Under the ROC Curve (AUC) (the higher the better, with a random classifier having an AUC of 0.5). 

\paragraph{Angular resolution}
The angular resolution ($\theta_{\sigma})$ is computed as the angle within which 68\% of the reconstructed gamma rays fall relative to their true direction. It is computed using 70\% of the gamma events with the highest gammaness. The lower the angular resolution, the better.

\paragraph{Energy resolution and bias}
The energy resolution ($E_{\sigma}$) and bias ($E_{\mu}$) are the energy relative errors respectively defined as one standard deviation and the median of the energy error between the gamma ray simulated and reconstructed energy. The lower the energy resolution, the better, while the closer the bias is to $0$, the better.

\paragraph{Summarized IRFs}
We propose to summarize the IRFs by integrating each figure over the energy. Although this procedure hides the energy-dependent performance of the model, it allows to illustrate and quantify the difference between each method with a more synthetic metric. More importantly, the IRFs are computed on both the source and the target datasets, while the integrated metrics presented in the following are only calculated on the target dataset. In addition, IACT performance changes greatly as a function of the energy, which is visible on the IRFs; thus it is important to refer to the IRFs in the online Zenodo record \cite{michael_dell_aiera_2024_13646001} for a complete understanding of the results.

\subsection{Training parameters}
The convergence of the models is ensured within 30 epochs. For the implementation of the DANN, the domain classifier follows the original paper and is composed of two fully-connected layers of 100 features. The weights of the model are updated using the Adam optimizer with a learning rate of $1e^{-3}$, and the weights associated UW (log-variances in that case) and GN are updated using SGD with a learning rate of $1e^{-3}$. Both domains are sampled with a batch size of 256 images, resulting in a mini-batch of size 512. In all cases, the optimizers have a weight decay of $1e^{-4}$. Overall, $\gamma$-PhysNet, DeepJDOT and DeepCORAL have $3.5M$ parameters, whereas the DANN based approach has $4.8M$ parameters. Such model configurations will be maintained throughout the remainder of the paper.

\subsection{Training procedure}
The $\gamma$-PhysNet is naturally trained including simulated gamma-ray and proton events in order to perform the particle classification. In addition, the multitask regression branches reconstruct the properties of the incident particle like the energy, direction, and impact point. The simulated protons are excluded from the loss calculation for those regression tasks, optimizing the performance of the full-event reconstruction towards gamma
rays.

\subsubsection{Impact of NSB}
In this section, our objective is to assess the influence of NSB on the $\gamma$-PhysNet. Simulations intrinsically contain a fixed amount of noise, following a Poisson distribution and applied independently to each pixel. The rate can be measured from the dataset by discarding the signal and keeping the background only. We measured a rate of $\lambda_{MC} = 1.77$. An additional Poisson noise is applied with parameter $\delta \lambda \in \{0.2, 0.4, 0.6, 0.8\}$, so that it is equivalent to follow a Poisson distribution of parameter $\lambda = \lambda_{MC} + \delta \lambda$. This considered range of $\delta \lambda$ represents a realistic scenario. In fact, the impact of the moonlight tends to provide greater values of $\delta \lambda$, typically $\delta \lambda > 0.5$. In this section, we train the $\gamma$-PhysNet on non-degraded images, which is equivalent to setting $\delta \lambda = 0$, and is referred to the best scenario in our previous work \cite{dellaiera2023uda}. Then, we evaluate the model on the specified range of noise rates.

The results are presented in Table \ref{tab:lst_nsb}. As expected, the $\gamma$-PhysNet is very sensitive to the difference in training and test distribution. In real-world scenarios, the level of NSB fluctuates over time due to external factors such as moonlight. The noise rate thus becomes a time-dependent function, $\lambda = \lambda(t)$, underlying the necessity for domain adaptation techniques.

\renewcommand{\arraystretch}{1.5} % 1.5 times the default row spacing
\begin{table}
\centering
\begin{tabular}{ c|p{1.02cm}p{1.02cm}p{1.02cm}p{1.02cm}p{1.02cm} }
$\delta \lambda$ & $0.00$ & $0.20$ & $0.40$ & $0.60$ & $0.80$ \\
\hline
\hline
$E_{\mu}$ & \textbf{0.02} & 0.10 & 0.44 & 5.06 & $>10^2$ \\
\hline
$E_{\sigma}$ & \textbf{0.27} & 0.31 & 1.02 & 38.0 & $>10^3$ \\
\hline
$\theta_{\sigma}$ & \textbf{0.24} & 0.26 & 0.41 & 1.27 & 1.98 \\
\hline
AUC & \textbf{0.87} & 0.83 & 0.69 & 0.54 & 0.50 \\
\end{tabular}
\vspace{0.2cm}
\caption{Impact of NSB on $\gamma$-PhysNet.}
\label{tab:lst_nsb}
\end{table}

\subsubsection{Sensitivity to label shifts}
To illustrate the impact of label shift on UDA, we train the DANN model as implemented in \cite{dellaiera2023uda} using multiple amount of gamma/proton ratios, denoted as  $r \in \{10^{-1}, 10^{-2}, 10^{-3}, 10^{-4}, 10^{-5}\}$, in the target dataset.

The results are presented in Table \ref{tab:lst_label_shift}. The performance metrics both on source and target are noticeably affected by the label shift, and the consequences are twofold. Firstly, the resolution and bias of energy and angle worsen as the ratio decreases, highlighting a loss in performance. Concurrently, the AUC on the target data increases, which is particularly visible at energy levels greater than $0.1$ TeV from the IRFs presented in the Zenodo record.

The impact of label shift on UDA, as demonstrated by the performance of the DANN model across different gamma/proton ratios, underscores the importance of addressing this problem to ensure robust and effective UDA.

\rev{However, Table \ref{tab:lst_label_shift} indicates that the performance remains relatively stable across several orders of magnitude in gamma fraction once the target domain becomes strongly proton dominated. In particular, the AUC remains nearly constant for ratios between $10^{-1}$ and $10^{-5}$, while the regression metrics vary only moderately. This suggests that the proposed conditioning mechanism is primarily sensitive to the presence of a strong label shift rather than to the exact value of the gamma/proton ratio. Consequently, moderate inaccuracies in the estimated target class proportions are not expected to significantly affect the adaptation process in realistic CTAO observations (where the gamma/proton ratio is always lower than $10^{-2}$).}

\renewcommand{\arraystretch}{1.5} % 1.5 times the default row spacing
\begin{table}
\centering
\begin{tabular}{ c|p{0.85cm}p{0.85cm}p{0.85cm}p{0.85cm}p{0.85cm}p{0.85cm} }
ratio & $10^{0}$ & $10^{-1}$ & $10^{-2}$ & $10^{-3}$ & $10^{-4}$ & $10^{-5}$ \\
\hline
\hline
$E_{\mu}$ & \textbf{0.06} & 0.10 & 0.10 & 0.10 & 0.12 & 0.14 \\
\hline
$E_{\sigma}$ & \textbf{0.30} & 0.32 & 0.32 & 0.33 & 0.35 & 0.35 \\
\hline
$\theta_{\sigma}$ & \textbf{0.26} & 0.28 & \textbf{0.26} & 0.27 & 0.29 & 0.29 \\
\hline
AUC & 0.82 & 0.81 & \textbf{0.83} & \textbf{0.83} & \textbf{0.83} & \textbf{0.83} \\
\end{tabular}
\vspace{0.2cm}
\caption{Impact of label shift on $\gamma$-PhysNet-DANN. Only the results on target data are shown. The ratio corresponds to the quantity $\dfrac{\# gammas}{\# protons}$ in the target dataset, while the source ratio is left unchanged ($\sim 1$).}
\label{tab:lst_label_shift}
\end{table}

\subsubsection{Multitask balancing}
We evaluate our UDA method combined with the selected MTB algorithms in the context of NSB of parameter $\delta \lambda=0.46$, in accordance to \cite{jacquemont2021cta}, and label shift of ratio $10^{-4}$ in the target set. 

The comparative study, available on the online repository, of our selected UDA techniques combined with GN across various values of the hyperparameter $\alpha \in \{0.1, 0.5, 1.5, 3.0\}$ reveals that DANN, DeepCORAL and DeepJDOT exhibit better performance with smaller $\alpha$ values, particularly when $\alpha = 0.1$. Specifically, DANN's efficacy noticeably diminishes when $\alpha > 0.1$, or sometimes fails to converge, as evidenced by outcomes on the source data. While DeepCORAL demonstrates less sensitivity to variations in this hyperparameter, examination of the IRFs indicates that the main distinction arises from the resolution at higher energy levels as $\alpha$ increases. Similarly, DeepJDOT corroborates DeepCORAL's conclusion, but higher values of $\alpha$ also deteriorate the performance on the source data.

A previous study \cite{dellaiera2023uda} highlighted the performance of the selected UDA algorithm combined with UW, but it did not consider the addition of label shift. In our investigations, the comparison of UW and GN reveals that all our methods have a better performance when paired with GN while DeepCORAL and DeepJDOT fails at converging with UW. This is potentially due to the fact that there are no guarantees that both methods meet the requirements of UW, and the uncertainty measure of the domain task cannot be compared to the variances of the physical reconstruction branches.

\subsubsection{Conditional domain adaptation}

\renewcommand{\arraystretch}{1.5} % 1.5 times the default row spacing
\begin{table}
\centering
\begin{tabular}{ c||p{0.8cm}p{1.8cm}|p{1.8cm}p{1.8cm}|p{1.8cm}p{1.8cm} }
 & \multicolumn{2}{c|}{DANN} & \multicolumn{2}{c|}{+C} & \multicolumn{2}{c}{+C+GL} \\
\hline
MTB & UW & GN & UW & GN & UW & GN \\
\hline
\hline
$E_{\mu}$ & 0.12 & 0.09 & 0.08 & 0.09 & \textbf{-0.06} & -0.12 \\
\hline
$E_{\sigma}$ & 0.35 & 0.33 & \textbf{0.31} & 0.33 & \textbf{0.31} & 0.32 \\
\hline
$\theta_{\sigma}$ & 0.29 & 0.27 & \textbf{0.26} & 0.27 & 0.31 & 0.31 \\
\hline
AUC & \textbf{0.83} & 0.82 & 0.82 & 0.82 & 0.72 & 0.77 \\
\end{tabular}
\caption{Ablation study of DANN. Only results on target data are indicated. +C specifies the conditioning, and +GL designates the use of the Gradient Layer.}
\label{tab:lst_dann}
\vspace{0.5cm}

\renewcommand{\arraystretch}{1.5} % 1.5 times the default row spacing
\centering
\begin{tabular}{ c||p{0.8cm}p{1.8cm}|p{1.8cm}p{1.8cm}|p{1.8cm}p{1.8cm} }
 & \multicolumn{2}{c|}{DeepCORAL} & \multicolumn{2}{c|}{+C} & \multicolumn{2}{c}{+C+GL} \\
\hline
MTB & UW & GN & UW & GN & UW & GN \\
\hline
\hline
$E_{\mu}$ & 0.20 & 0.19 & 0.18 & \textbf{0.13} & 0.18 & 0.17 \\
\hline
$E_{\sigma}$ & 0.44 & 0.40 & 1.12 & \textbf{0.33} & 1.13 & 0.37 \\
\hline
$\theta_{\sigma}$ & 0.33 & \textbf{0.25} & 0.45 & 0.26 & 0.45 & 0.27 \\
\hline
AUC & \textbf{0.82} & \textbf{0.82} & 0.54 & \textbf{0.82} & 0.59 & \textbf{0.82} \\
\end{tabular}
\caption{Ablation study of DeepCORAL. Only results on target data are indicated. +C specifies the conditioning, and +GL designates the use of the Gradient Layer.}
\label{tab:lst_deepcoral}
\vspace{0.5cm}

\renewcommand{\arraystretch}{1.5} % 1.5 times the default row spacing
\centering
\begin{tabular}{ c||p{0.8cm}p{1.8cm}|p{1.8cm}p{1.8cm}|p{1.8cm}p{1.8cm} }
 & \multicolumn{2}{c|}{DeepJDOT} & \multicolumn{2}{c|}{+C} & \multicolumn{2}{c}{+C+GL} \\
\hline
MTB & UW & GN & UW & GN & UW & GN \\
\hline
\hline
$E_{\mu}$ & 0.55 & 0.13 & -0.14 & \textbf{0.11} & -0.23 & 0.19 \\
\hline
$E_{\sigma}$ & 0.86 & \textbf{0.37} & 0.89 & \textbf{0.37} & 0.74 & 0.41 \\
\hline
$\theta_{\sigma}$ & 0.34 & 0.29 & 0.45 & 0.29 & 0.46 & \textbf{0.28} \\
\hline
AUC & 0.54 & \textbf{0.81} & 0.57 & 0.80 & 0.51 & \textbf{0.81} \\
\end{tabular}
\caption{Ablation study of DeepJDOT. Only results on target data are indicated. +C specifies the conditioning, and +GL designates the use of the Gradient Layer.}
\label{tab:lst_deepjdot}
\end{table}

In this section, our focus lies in quantifying the benefits of conditional domain adaptation. The results are presented in Table \ref{tab:lst_dann}, Table \ref{tab:lst_deepcoral} and Table \ref{tab:lst_deepjdot}, and respectively describe DANN, DeepCORAL and DeepJDOT in 3 contexts: the initial algorithm (see Eqs \ref{eq:loss_deepcoral},\ref{eq:loss_deepjdot},\ref{eq:loss_dann}), the conditioning (named +C; see Eqs \ref{eq:loss_cdann},\ref{eq:loss_cdeepjdot},\ref{eq:loss_cdeepcoral}), and the inclusion of GL (named +GL; see Eq \ref{eq:gradient_layer}). Firstly, the extension of the $\gamma$-PhysNet with CDANN shows that, while GN initially outperformed UW without conditioning, we finally observe a change in performance dynamics. CDANN combined with UW yields the best overall results for the energy bias and resolution, demonstrating that the introduction of conditioning allows to recover from label shift, ultimately leading to a restoration of performance levels comparable to those observed previously in \cite{dellaiera2023uda}. Furthermore, CDeepCORAL combined with GN results in a significant improvements in energy bias and resolution at lower energy levels compared to DeepCORAL, although a slight reduction in resolution is observed at intermediate levels. These improvements are consistent across both the source and target domains. However, the conditioning with UW degrades the energy resolution. Finally, CDeepJDOT paired with GN reveals a slight increase in performance on the target energy bias, and the overall metrics exhibit a deterioration while paired with UW. The addition of conditional domain adaptation does not allow for DeepJDOT and DeepCORAL to converge while using UW.

\subsubsection{Inclusion of gradient weighting}
We evaluate the benefits of the GL in the UDA models. In the case of DANN, weighting the gradients deteriorate the performance whether it is associated with UW or GN. Yet, in the case of DeepCORAL and DeepJDOT, the GL combined with GN significantly improve the performance on the source dataset. However, the outcomes on the target events are more nuanced. While metrics indicate improvements at higher energies, the lower part of the range exhibits inferior results. Moreover, pairing GL with UW in both cases fails to rectify the convergence issues.

\rev{While the proposed GL acts as a lightweight scheduler to mitigate early-stage gradient conflicts, it does not explicitly resolve the geometric opposition of the task gradients. As observed in the training dynamics, particularly for DeepJDOT, conflicting gradients in both direction and amplitude remain a challenge. Future work should explore advanced gradient-surgery approaches, such as PCGrad \cite{yu2020pcgrad} or conflict-aware optimization, which project conflicting gradients onto normal planes. Implementing such techniques could further stabilize the joint optimization of the domain confusion and physical reconstruction tasks, albeit at a higher computational cost.}

\section{Conclusion}
\label{sec:conclusion}
In this paper, we combined and analysed the integration of UDA into MTB, revealing comparable or improved performances while reducing the computation cost of the loss weighting optimization. Moreover, the implementation of the GL, that acts as a scheduler for the domain adaptation task, can be beneficial for methods that output a strong and conflicting gradients. Furthermore, we introduced conditional domain adaptation to take into account the strong label shift that arises from LST observations, improving the performance of our methods. 

\rev{Moreover, although DeepJDOT provides a mathematically robust framework for domain alignment via OT, it presents significant practical challenges for large-scale astrophysical datasets. The computation of the pairwise cost matrix and the OT plan scales quadratically, imposing strict upper limits on batch sizes due to memory and computational time constraints. Furthermore, as observed in our experiments, DeepJDOT is highly sensitive to hyperparameter tuning and exhibits massive initial gradient norms that destabilize multitask convergence. Due to these training instabilities, DeepJDOT is less suited for the noisy, large-scale reality of real telescope observations, making other approaches a much more stable and scalable choice for CTAO data analysis.}
\rev{In conclusion, o}ur comparative study, conducted on simulations with multiple NSB and label shift, leads to the selection of CDANN paired with UW as the most promising tool for the detection and analysis of real gamma-ray sources. Results on real telescope data will be presented in a subsequent work.

\rev{This work intentionally focused exclusively on simulations to control the impact of data degradations while retaining the ground truth labels. Additional discrepancies may be present in real observations, and the impact of UDA methods on LST-1 data will be studied in a subsequent work.
}

\bibliographystyle{elsarticle-num}
\bibliography{refs}

\end{document}